\documentclass[aps,pra,showpacs,onecolumn]{revtex4}

\usepackage{latexsym,makeidx}
\usepackage{amsmath,amsfonts,amssymb,amsbsy,amscd}
\usepackage[dvips]{graphicx}
\usepackage{bbm}

\newcommand{\Tr}{\ensuremath{\textrm{Tr}}}

\newcommand{\id}{\ensuremath{\mathbbm{1}}}

\newcommand{\bbi}{\ensuremath{\mathbbm{i}}}
\newcommand{\bra}[1]{\ensuremath{\langle#1|}}
\newcommand{\ket}[1]{\ensuremath{|#1\rangle}}
\newcommand{\braket}[2]{\ensuremath{\langle#1|#2\rangle}}

\newcommand{\mcN}{\ensuremath{\mathcal{N}}}
\newcommand{\mcJ}{\ensuremath{\mathcal{J}}}
\newcommand{\mcA}{\ensuremath{\mathcal{A}}}

\newcommand{\mcH}{\ensuremath{\mathcal{H}}}
\newcommand{\brac}[1]{\ensuremath{\left(#1\right)}}

\begin{document}
\title{Approximate quantum data storage and teleportation}
\author{Thomas Laustsen and Klaus M\o lmer}
\affiliation{QUANTOP - Danish Quantum Optics Center\\
  Institute of Physics and Astronomy, University of Aarhus\\
  DK-8000 \AA rhus C, Denmark}
\date{\today}
\begin{abstract}
In this paper we present an optimal protocol by which an unknown state on a Hilbert space of dimension $N$ can be approximately stored in an $M$-dimensional quantum  system or be approximately teleported via an $M$-dimensional quantum channel. The fidelity of our procedure is determined for pure states as well as for mixed states and states which are entangled with auxiliary quantum systems of varying Hilbert space dimension, and it is compared with theoretical results for the maximally achievable fidelity.
\end{abstract}

\maketitle

\section{Introduction}

Imagine the following scenario: We are given an unknown quantum state of an $N$-level system, and we want to store that state, but as storage medium we have only a classical storage device and a physical $M$-level quantum system ($M<N$). What is the optimal protocol for this task, and with which probability will we be able to retrieve the original state after the process? This problem is formally equivalent to the one of transporting an unknown state, but having as transport medium only a classical channel and an $M$-level quantum channel, either in the form of a portable quantum system with $M$ levels or a teleportation channel with an initially prepared entangled state of the form $\frac{1}{\sqrt{M}}\sum_{i=1}^{M}|i_A\rangle|i_B\rangle$.

In the present paper we shall present a protocol to achieve these goals with a mean fidelity (to be defined below) for a pure input state of $F=(M+1)/(N+1)$. And we shall show that the $N$-dimensional component of an entangled state of an $N$- and an $R$-dimensional system can be stored or transported by an $M$-dimensional system, so that the entangled state can be reconstructed (now, possibly with the two components spatially separated), with a mean fidelity of $F=(MR+1)/(NR+1)$. Teleportation of mixed states of rank $R\le N$, can be done with a Bures fidelity given by the same expression.

The problem is closely linked to the issue of entanglement manipulation and quantum state transformation, and some of the above mentioned results can indeed be tested against special cases of the maximum fidelity of faithful transformation of a pure entangled state into a maximally entangled state of two $N$-level systems as computed by Vidal \emph{et al.} \cite{art80}, Horodeckis \cite{art65}. The cited works identify the optimum theoretical transformation of the quantum channel, whereas our approach offers a different perspective as it deals with explicit operations on the incident quantum state. Related work has also been published by Banaszek in Ref.~\cite{art157}. Here the optimal fidelity for teleportation of pure states was found for an arbitrary pure state channel, and the result was found to agree with that of Ref.~\cite{art80}. We will return to a discussion of the differences between our proposal and Banaszek's work in the discussion.

We shall formulate our problem as the one of teleportation of an $N$-dimensional state through a perfect $M$-dimensional quantum channel. In Sec.~\ref{sec:cutting-procedure} we shall present our very simple scheme, which has the same fidelity, although it differs from the ones of Ref.~\cite{art65} and Ref.~\cite{art157}. In Sec.~\ref{sec:entangled-states} we compute the fidelity of teleportation of entangled states and in Sec.~\ref{sec:mixed-states} we consider the problem of general mixed input states of the quantum system. In Sec.~\ref{sec:discussion} we summarize our conclusions and discuss implications of our results. 


\section{The Cutting Procedure}
\label{sec:cutting-procedure}

Assume that two parties, Alice and Bob, share a maximally entangled $M \times M$-dimensional state (the channel),
\begin{equation}
\label{eq:channel}
\ket{\Psi_M}=\frac{1}{\sqrt{M}}\sum_{i=1}^{M}\ket{i_A,i_B},
\end{equation}
and that Alice possesses an arbitrary, unknown $N$-dimensional pure state that she wants to transfer to a quantum system located at Bob's place with the greatest accuracy possible using only local quantum operations and classical communication. Transferring an $N$-dimensional quantum state $\ket{\psi}$ through an $M$-dimensional quantum channel cannot be done with unit fidelity if $M<N$ \cite{art15}, but many different methods can be applied to do it approximately. What is the best teleportation scheme and what is the corresponding fidelity of the state which Bob recieves? 

The method we are going to use is to first reduce the dimensionality of the state from $N$ to $M$ by a positive operator-valued measurement (POVM) \cite{Peres}, and by subsequently teleporting the resulting state perfectly through our $M$-dimensional quantum channel. If we choose the set $\{\ket{\phi_j}\}_{j=1}^{N}$ to form an orthonormal basis for the $N$-dimensional Hilbert-space of the initial state $\ket{\psi}$, the set of operators
\begin{equation}
\label{eq:A}
\{\hat{A_{\bbi}}\},\quad 
\hat{A_{\bbi}}=\frac{1}{\mcN}\sum_{j=1}^{M}\ket{\phi_{i_j}}\bra{\phi_{i_j}}
\end{equation}
constitutes our POVM that will be used to perform the $N\to M$ cut. The constant $\mcN=\binom{N-1}{M-1}$ can be determined from the normalisation condition, $\sum_{\bbi}\hat{A_{\bbi}}=\id$, and the sets of numbers $\bbi=\{i_1,\cdots,i_M\}$ runs through all the $\binom{N}{M}$ possible choices. 

\subsection{Pure States}
\label{sec:pure-states}
We first consider the case of a pure initial state. The measurement outcome corresponding to $\hat{A_{\bbi}}$ occurs with probability $p_{\bbi}=\bra{\psi}\hat{A_{\bbi}}\ket{\psi}$ in which case the projected state is $\ket{\tilde{\psi}_{\bbi}}=\frac{\hat{A_{\bbi}}\ket{\psi}}{\sqrt{\bra{\psi}\hat{A_{\bbi}}^{\dagger}\hat{A_{\bbi}}\ket{\psi}}}$. The fidelity of $\ket{\tilde{\psi}_{\bbi}}$ with respect to $\ket{\psi}$ is just the overlap 
\begin{equation}
  \label{eq:49}
  f_{\bbi}=|\braket{\psi}{\tilde{\psi}_{\bbi}}|^2,
\end{equation}
which we see is also equal to $f_{\bbi}=\mcN \bra{\psi}\hat{A_{\bbi}}\ket{\psi}$. The average fidelity (averaged over measurement outcomes and over incident states) is therefore
\begin{equation}
\label{cut_formula} 
\begin{split}
F_{N \to M} &=\frac{1}{\mcA_N}\int d\Omega_{N} \sum_{\bbi}\mcN\bra{\psi}\hat{A_{\bbi}}\ket{\psi}^2\\
&= \frac{1}{\mcA_N}\int d\Omega_{N} \sum_{\bbi} \frac{1}{\mcN } \brac{\sum_{j=1}^{M} |\braket{\phi_{i_j}}{\psi}|^2}^2\\
&=\frac{1}{\mcA_N}\int d\Omega_{N}\frac{1}{\mcN}\left[\mcN\frac{M-1}{N-1}\brac{\sum_{j=1}^{N}|\braket{\phi_j}{\psi}|^2}^2+\brac{\mcN-\mcN\frac{M-1}{N-1}}\sum_{j=1}^{N}|\braket{\phi_{j}}{\psi}|^4\right]\\
&=\frac{M-1}{N-1}+\frac{N-M}{N-1}\frac{1}{\mcA_N}\int d\Omega_{N}\sum_{j=1}^{N} |\braket{\phi_{j}}{\psi}|^4
,
\end{split}
\end{equation}
where $d\Omega_{N}$ is the appropriate ``surface area''-element on the unit hypersphere in the $N$-dimensional complex Hilbert space, and $\mcA_N\equiv\int d\Omega_{N}$. Since we average over input states, we do not need to average over different choices of the orthogonal basis. Equation \eqref{cut_formula} is therefore independent of the choice of basis states $\{\ket{\phi_j}\}$.

The integral in the last line in equation~(\ref{cut_formula}) we recognize as the average fidelity of estimating a state after a von Neumann measurement \cite{Peres} in the basis $\{\ket{\phi_j}\}$
\begin{equation}
  \label{eq:4}
  F_{N\to 1}=\frac{1}{\mcA_N}\int d\Omega_{N}\sum_{j=1}^{N} |\braket{\phi_{j}}{\psi}|^4,
\end{equation}
and we thus obtain the nice relation
\begin{equation}
  \label{eq:16}
  F_{N\to M}=\frac{M-1}{N-1}+\frac{N-M}{N-1}F_{N\to 1}.
\end{equation}
The problem is now reduced to that of calculating $F_{N\to 1}$, which is done in the following way. First we simplify equation~(\ref{eq:4}) to
\begin{equation}
  \label{eq:1}
  F_{N\to 1}=\frac{1}{\mcA_N}N\int d\Omega_{N}|\braket{\phi_{1}}{\psi}|^4,
\end{equation}
by noting that all $N$ components of the state $\ket{\psi}=\sum_{j=1}^{N}\braket{\phi_{j}}{\psi} \ket{\phi_{j}}$ will contribute equally to the sum after the averaging over states.

As a general representation for a state on the unit hypersphere in $\mathbb{C}^N$ we choose
\begin{equation}
  \label{eq:2}
  \ket{\psi}=\brac{\begin{array}{l}
       \cos\theta_{1}e^{i\phi_{1}}\\
       \sin\theta_{1}\cos\theta_{2}e^{i\phi_{2}}\\
       \sin\theta_{1}\sin\theta_{2}\cos\theta_{3}e^{i\phi_{3}}\\
       \hspace{.3cm}\vdots\hspace{1cm}\vdots\hspace{1cm}\ddots\hspace{.5cm}\ddots\\
       \sin\theta_{1}\sin\theta_{2}\cdots\sin\theta_{N-2}\cos\theta_{N-1}e^{i\phi_{N-1}}\\
       \sin\theta_{1}\sin\theta_{2}\cdots\sin\theta_{N-2}\sin\theta_{N-1}e^{i\phi_{N}}
    \end{array}},
\hspace{.5cm}
\begin{array}{c}
0\le\theta_{1},\ldots,\theta_{N-1}\le\frac{\pi}{2}\\
0\le\phi_{1},\ldots,\phi_{N}\le 2\pi
\end{array}
,
\end{equation}
and the corresponding measure, $d\Omega_{N}$, is found in the appendix to be
\begin{equation}
  \label{eq:5}
  d\Omega_{N}=\prod_{k=1}^{N-1}\brac{\cos\theta_{k}\sin\theta_{k}\brac{\sin^2\theta_{k}}^{N-k-1}d\theta_{k}d\phi_{k}}d\phi_{N},
\end{equation}
The integral~(\ref{eq:1}) can now be evaluated as
\begin{equation}
  \label{eq:7}
\begin{split}
    F_{N\to 1}&=\frac{N}{\mcA_N}\int d\Omega_{N}|\braket{\phi_{1}}{\psi}|^4
=N\frac{\int d\Omega_{N}\cos^4\theta_{1}}{\int d\Omega_{N}}\\
&=N\frac{\int_{0}^{\frac{\pi}{2}}\cos\theta_{1}\sin\theta_{1}\brac{\sin^2\theta_{1}}^{N-2}\cos^4\theta_{1}d\theta_{1}}{\int_{0}^{\frac{\pi}{2}}\cos\theta_{1}\sin\theta_{1}\brac{\sin^2\theta_{1}}^{N-2}d\theta_{1}}\\
&=\frac{2}{N+1},
\end{split}
\end{equation}
and inserting this into the formula (\ref{eq:16}) yields the result
\begin{equation}
  \label{eq:8}
F_{N \to M}=\frac{M+1}{N+1}.
\end{equation}

This value for the fidelity is in agreement with the following result from the literature \cite{art65}
\begin{equation}
  \label{eq:17}
  F_{N\to M}^{(\textrm{opt})}=\frac{Nf_s(\Psi_M)+1}{N+1}
\end{equation}
where $f_s(\Psi_M)=M/N$ is the \emph{singlet fraction} of the channel, i.e. the fidelity by which the $M$-dimensional channel (\ref{eq:channel}) can be transformed into an $N$-dimensional one (with which perfect teleportation can be subsequently achieved for any state in the N-dimensional Hilbert space). This agreement is reassuring, since for the task of teleportation we deal with the same shared quantum resources. By cutting the system to fit the resources rather than by extending the resources to fit the system, the present approach presents an alternative analysis to Ref.~\cite{art65}, and it treats simultaneously the tasks of teleportation and of storage or physical transport of a quantum state. The value is also in agreement with the result of Banaszek \cite{art157}, as we will get back to in the discussion.

The explicit calculation of the fidelity based on wave function overlaps also lends itself to further analysis, as we shall turn to in the following sections and in the discussion.


\section{Entangled States}
\label{sec:entangled-states}

We now consider a pure state $\ket{\psi}\in\mcH\otimes\mcH_R$ shared between Alice and R of which we want to teleport Alice's part to Bob without acting on the R degrees of freedom. As in the previous section we assume to have the channel (\ref{eq:channel}) and we apply the same protocol. 

Thus we need to calculate the fidelity $f_{\bbi}$ of the teleported state 
$\ket{\tilde{\psi}_{\bbi}}=
\frac{(\hat{A_{\bbi}}\otimes\id_R)\ket{\psi}}{\sqrt{\bra{\psi}
\brac{\hat{A_{\bbi}}^{\dagger}\hat{A_{\bbi}}\otimes\id}\ket{\psi}}}$ w.r.t. the initial state $\ket{\psi}$,
\begin{equation}
  \label{eq:19}
  f_{\bbi}=\frac{|\bra{\psi}(\hat{A_{\bbi}}\otimes \id)\ket{\psi}|^2}{\bra{\psi}\brac{\hat{A_{\bbi}}^{\dagger}\hat{A_{\bbi}}\otimes\id}\ket{\psi}}.
\end{equation}
The probability of the measurement outcome corresponding to $\hat{A_{\bbi}}$ is \begin{equation}
  \label{eq:27}
  p_{\bbi}=\bra{\psi}(\hat{A_{\bbi}}\otimes\id_R)\ket{\psi},
\end{equation}
and the expression for $f_{\bbi}$ simplifies to $f_{\bbi}=\frac{p_{\bbi}^{2}}{p_{\bbi}/\mcN}=\mcN p_{\bbi}$. Hence the teleportation fidelity is
\begin{equation}
  \label{eq:6}
F_{N\to M}=\frac{1}{\mcA_{NR}}\int d\Omega_{NR}\sum_{\bbi} p_{\bbi}f_{\bbi}
=\frac{1}{\mcA_{NR}}\int d\Omega_{NR}\mcN\sum_{\bbi} \bra{\psi}(\hat{A_{\bbi}}\otimes\id_R)\ket{\psi}^2,
\end{equation}
where $NR=\dim\brac{\mcH\otimes\mcH_R}$. Inserting the expression (\ref{eq:A}) for the $\hat{A_{\bbi}}$'s we find
\begin{equation}
  \label{eq:30}
  \begin{split}
  F_{N\to M}&=\frac{1}{\mcA_{NR}}\int d\Omega_{NR}\frac{1}{\mcN}\sum_{\bbi}\brac{\sum_{j=1}^{M}\brac{\sum_{k=1}^{R} |\braket{\phi_{i_{j}},k}{\psi}|^2}}^2\\
  &=\frac{1}{\mcA_{NR}}\int d\Omega_{NR}\frac{1}{\mcN}\left[
\mcN\frac{M-1}{N-1}\brac{\sum_{j=1}^{N}\sum_{k=1}^{R} |\braket{\phi_{j},k}{\psi}|^2}^2+\brac{\mcN-\mcN\frac{M-1}{N-1}}\sum_{j=1}^{N}\brac{\sum_{k=1}^{R} |\braket{\phi_{j},k}{\psi}|^2}^2\right]\\
  &=\frac{M-1}{N-1}+\frac{N-M}{N-1}F_{N\to 1}
  \end{split}
\end{equation}
as in the pure state case, equation~(\ref{eq:16}).

In evaluating $F_{N\to 1}$ we can again use the isotropy of the state space to get rid of a sum,
\begin{equation}
  \label{eq:31}
  F_{N\to 1}=\frac{1}{\mcA_{NR}}\int d\Omega_{NR}\sum_{j=1}^{N}\brac{\sum_{k=1}^{R} |\braket{\phi_{j},k}{\psi}|^2}^2
  =N\frac{1}{\mcA_{NR}}\int d\Omega_{NR}\brac{\sum_{k=1}^{R} |\braket{\phi_{1},k}{\psi}|^2}^2.
\end{equation}

Carrying out the square operation and applying the isotropy property and the representation (\ref{eq:2}) again, this expression reduces to
\begin{equation}
  \label{eq:33}
  \begin{split}
  F_{N\to 1}&=\frac{N}{\mcA_{NR}}\int d\Omega_{NR}\brac{\sum_{k=1}^{R} |\braket{\phi_{1},k}{\psi}|^4+\sum_{k<k'} 2|\braket{\phi_{1},k}{\psi}|^2|\braket{\phi_{1},k'}{\psi}|^2}\\
  &=\frac{N}{\mcA_{NR}}\int d\Omega_{NR}\brac{R|\braket{\phi_{1},1}{\psi}|^4+\frac{R(R-1)}{2} 2|\braket{\phi_{1},1}{\psi}|^2|\braket{\phi_{1},2}{\psi}|^2}\\
  &=\frac{NR}{\mcA_{NR}}\int d\Omega_{NR}\brac{\cos^4\theta_1+(R-1)\cos^2\theta_1\sin^2\theta_1\cos^2\theta_2},
  \end{split}
\end{equation}
which again can be evaluated using the measure (\ref{eq:5}), and the result is
\begin{equation}
  \label{eq:34}
  F_{N\to 1}=\frac{R+1}{NR+1}.
\end{equation}
Inserting this into the formula (\ref{eq:30}) now also yields the entangled state teleportation fidelity,
\begin{equation}
  \label{eq:35}
  F_{N\to M}=\frac{MR+1}{NR+1}.
\end{equation}

We observe the interesting result that this expression depends on the dimension of the space of the auxiliary system. The fidelity for transmitting 
or storing the $N$-dimensional component of an entangled state through an
$M$-dimensional channel is almost a factor $R$ times larger than
the fidelity if the entire entangled state (of dimension $NR$) 
should be compressed to the channel size. This is because 
we only swap the entanglement from Alice and the auxiliary 
system to Bob and the auxiliary system, and the 
auxiliary component is not acted upon by the protocol. 
This entanglement swapping is perfect, when $N=M$.

\section{Mixed States}
\label{sec:mixed-states}

The case of mixed states requires a special treatment. The appropriate measure of the fidelity of $\tilde{\rho}$ w.r.t. $\rho$ is the \emph{Bures fidelity} or \emph{Uhlmann transition probability}, see e.g. \cite{art58,art253}
\begin{equation}
  \label{eq:20}
  F(\rho,\tilde{\rho})
  =\brac{\Tr\brac{\sqrt{\rho}\tilde{\rho}\sqrt{\rho}}^{1/2}}^2,
  \end{equation}
which can also be written
\begin{equation}
  \label{eq:18}
  F(\rho,\tilde{\rho}) =\max |\braket{\phi}{\tilde{\phi}}|^2,
\end{equation}
where the maximum is taken over all possible \emph{purifications}, $\ket{\phi}$ and $\ket{\tilde{\phi}}$, of $\rho$ and $\tilde{\rho}$ respectively. By a purification of a mixed state $\rho$ acting on $\mcH$ is meant a pure state $\ket{\xi}\in\mcH\otimes\mcH_R$ fulfilling the condition $\rho=\Tr_{R}\ket{\xi}\bra{\xi}$. If we write the Schmidt decomposition \cite{Peres} of $\ket{\xi}$
\begin{equation}
  \label{eq:21}
  \ket{\xi}=\sum_i \sqrt{\lambda_i}\ket{i}\otimes\ket{i_R}
\end{equation}
we see that all the different purifications of $\rho$ correspond to different choices of orthonormal basis sets $\{i_R\}$ for $\mcH_R$ (the $\lambda_i$'s, i.e. the eigenvalues of $\rho$, are the same in all purifications). Since these are related by a unitary transformation, any purification can be found from a particular one by a transformation $\ket{\xi}\to(\id\otimes U)\ket{\xi}$, where $U$ is unitary. Thus, if $\ket{\phi_0}$ and $\ket{\tilde{\phi}_0}$ are two particular purifications of $\rho$ and $\tilde{\rho}$, the fidelity is
\begin{equation}
  \label{eq:22}
  F(\rho,\tilde{\rho})=\max |\braket{\phi}{\tilde{\phi}}|^2
  =\max_U |\bra{\phi_0}(\id\otimes U)\ket{\tilde{\phi}_0}|^2,
\end{equation}
where the maximum is now over unitary transformations $U$ on $\mcH_R$.

In the present situation we would like to teleport the general mixed state
\begin{equation}
  \label{eq:23}
  \rho=\sum_{ij}\rho_{ij}\ket{\phi_i}\bra{\phi_j}
\end{equation}
through the channel (\ref{eq:channel}) with the above protocol. Thus, in analogy to the previous section, we need to calculate the fidelity $f_{\bbi}$ of the teleported state $\tilde{\rho}_{\bbi}=\frac{\hat{A}_{\bbi}^{}\rho \hat{A}_{\bbi}^{\dagger}}{\Tr(\hat{A}_{\bbi}^{}\rho \hat{A}_{\bbi}^{\dagger})}$ w.r.t. the initial state $\rho$. We choose an arbitrary purification $\ket{\psi}$ of $\rho$ and from this we construct a possible purification $\ket{\tilde{\psi}_{\bbi}}=\brac{\id\otimes U}\frac{(\hat{A_{\bbi}}\otimes\id_R)}{\sqrt{\Tr(\hat{A}_{\bbi}^{}\rho \hat{A}_{\bbi}^{\dagger})}}\ket{\psi}$ of $\tilde{\rho}_{\bbi}$. Inserting these into equation~(\ref{eq:22}) the particular fidelity is seen to be given by 
\begin{equation}
  \label{eq:24}
  f_{\bbi}=\max_U \frac{|\bra{\psi}(\hat{A_{\bbi}}\otimes U)\ket{\psi}|^2}{\Tr(\hat{A}_{\bbi}^{}\rho \hat{A}_{\bbi}^{\dagger})}.
\end{equation}
Letting
\begin{equation}
  \label{eq:25}
  \ket{\psi}=\sum_{j=1}^{N}\sum_{k}c_{jk}^{}\ket{\phi_j,k},
\end{equation}
where $\ket{\phi_j,k}$ is the tensor product of $\ket{\phi_j}$ in $\mcH$ and the $k^{\textrm{th}}$ basis vector in $\mcH_{R}$, and using the expression (\ref{eq:A}) for the $\hat{A_{\bbi}}$'s this reduces to
\begin{equation}
  \label{eq:26}
  \begin{split}
  f_{\bbi}&=\frac{1}{\Tr(\hat{A}_{\bbi}^{}\rho \hat{A}_{\bbi}^{\dagger})}\max_U \left|\sum_{jj'}\sum_{kk'}c_{jk}^{*}c_{j'k'}^{}\bra{\phi_j}\hat{A_{\bbi}}\ket{\phi_{j'}}\bra{k}U\ket{k'}\right|^2\\
  &=\frac{1}{\Tr(\hat{A}_{\bbi}^{}\rho \hat{A}_{\bbi}^{\dagger})} \max_U \left|\frac{1}{\mcN}\sum_{j=1}^{M}\brac{\sum_k c_{i_{j}k}^{*}\bra{k}}U\brac{\sum_k c_{i_{j}k}^{}\ket{k}}\right|^2\\
  &=\frac{|\bra{\psi}(\hat{A_{\bbi}}\otimes \id)\ket{\psi}|^2}{\Tr(\hat{A}_{\bbi}^{}\rho \hat{A}_{\bbi}^{\dagger})}
\end{split}
\end{equation}
\label{sec:mixed-states-1}
where, in the second step, we see that the fact that the optimum is taken at $U=\id$ implies that $f_{\bbi}$ is given by the same expression as in the previous section. 

For each purification $\ket{\psi}$ of $\rho$ we have now the expressions (\ref{eq:30}),~(\ref{eq:31}) for the fidelity. We wish to perform the averaging over input states in the most uniform manner. In the mixed state case, however, this is an ambiguous task, as discussed by \.{Z}yczkowski \emph{et al.} \cite{art291} and references therein. Here we make the natural choice of averaging over pure states $\ket{\psi}\in\mcH\otimes\mcH_R$ from which density matrices on $\mcH$ are given by the mapping $\ket{\psi}\longrightarrow\Tr_R\ket{\psi}\bra{\psi}$. Thus we have from (\ref{eq:31}) 
\begin{equation}
  \label{eq:32}
  F_{N\to 1}=N\frac{1}{\mcA_{NR}}\int d\Omega_{NR}\brac{\sum_{k=1}^{R} |\braket{\phi_{1},k}{\psi}|^2}^2=\frac{R+1}{NR+1}
\end{equation}
as in the previous section, and hence the result (\ref{eq:35}) holds equally well in this situation. Note that the averaging procedure introduces the dependence on the dimension $R$ of the auxiliary space corresponding to different probability measures $P_{N,R}(\rho)$ on the space of mixed states discussed in Ref.~\cite{art291}, where also other measures are discussed.

Concerning the optimality of our protocol, we refer to Bru{\ss} 
\emph{et al.} \cite{art168} for state estimation 
and to the Horodeckis \cite{art65} for teleportation 
in the pure state case, treated in Sec.~\ref{sec:pure-states}. 
Concerning teleportation of pure 
entangled states, treated in Sec.~\ref{sec:entangled-states}, 
we note that the auxiliary component (of dimension $R$) of the 
entangled system is the same in the initial and final state. 
This component is not acted upon by our protocol, {\it i.e.}, 
we can look at this part of the entangled state as if it is 
transferred with unit fidelity to the final state through a 
maximally entangled channel 
$\frac{1}{\sqrt{R}}\sum_{k=1}^{R}\ket{k,k} \in \mcH_R\otimes\mcH_R$.
The channel employed in our protocol results in the same final
state as a teleportation channel based on a maximally 
entangled $MR$ - dimensional entangled state, and thus the optimality 
of our protocol follows from the optimality in the pure state case, 
proved in Ref.~\cite{art65}.

For the case of a general mixed state, the optimal protocol is 
linked to what specific distribution we choose. The optimum in 
Eq.~(\ref{eq:26}) is taken for $U=\id$ for any POVM for which we 
can find a common basis in which all of its operators 
$\hat{A_{\bbi}}$ are diagonal, and with the 
chosen averaging procedure our protocol is optimal among such POVM's. 
We have not succeded in constructing a proof that one cannot perform 
better with POVM's that are not on this form. We find it most likely, 
however, that our protocol is optimal.

\section{Discussion}
\label{sec:discussion}

To summarize we have found a specific protocol with which the optimal fidelity is reached for teleportation of an $N$-dimensional state through an $M$-dimensional quantum channel or for storage in an $M$-dimensional system.

\begin{figure}
  \centering
  \includegraphics[width=7cm]{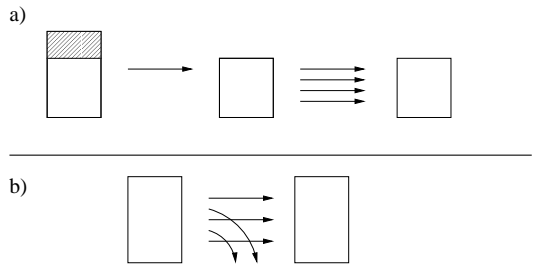}
  \caption{Schematic drawings of a) Our scheme, and b) The schemes of \cite{art65} and \cite{art157}. The single arrow in a) indicates the ``cut''-procedure of the present paper, and the four straight arrows indicate perfect teleportation, as in \cite{art15}. The straight and bent arrows in b) indicate the teleportation with a nonperfect channel, applied by Ref.'s~\cite{art65,art157}.}
  \label{fig:cut}
\end{figure}

As mentioned in the introduction related work exists in the literature. Figure~\ref{fig:cut} illustrates the difference between our protocol and the ones of Ref.~\cite{art65} and Ref.~\cite{art157}. Here the only disturbance of the state, namely the ``cut'', happens while it is still located at Alice's place, and before it is brought into contact with the teleportation channel. As a result of this the state is stored at Alices place in a state of size sufficiently small to be teleported perfectly (the four straight arrows) to Bob's place using the standard protocol for teleportation of $M$-dimensional states \cite{art15}. 

In the protocols of Ref.'s~\cite{art65} and \cite{art157,art15} the channel is assumed to be of dimension $N\times N$, and the disturbance of the state happens during the teleportation procedure (fig.~\ref{fig:cut}b). As discussed in these references, this offers a straightforward generalization to channels which are not on the form of Eq.~(\ref{eq:channel}). For the specific problem considered in this paper, however, our protocol has some advantages and offers a different perspective on the problem.

Banaszek \cite{art157} also discusses how much information is revealed about the state by the measurement protocol: This is quantified by the fidelity, $\bar{f}_{st.est.}$ defined as the average overlap between the actual state and our best guess based on the measurement. With our protocol, we can achieve the upper bound found by Banaszek: With probability $p_{\bbi}=\bra{\psi}\hat{A_{\bbi}}\ket{\psi}$ the outcome is $\bbi$, and our best guess on the state $\ket{\psi}$ will be $\ket{\psi_{\bbi}^{(\textrm{guess})}}=\ket{\phi_{i_1}}$. Thus we find the particular fidelity to be $f_{st.est.,\bbi}=\left|\braket{\phi_{i_1}}{\psi}\right|^2$, and hence
\begin{equation}
  \label{eq:29}
  \begin{split}
    \bar{f}_{st.est.}&=\frac{1}{\mcA_N}\int d\Omega_{N}\sum_{\bbi} p_{\bbi}f_{st.est.,\bbi}\\
    &=\frac{1}{\mcA_N}\int d\Omega_{N}\frac{1}{\mcN}\sum_{\bbi}\left|\braket{\psi}{\phi_{i_1}}\right|^2\sum_{j=1}^{M}\left|\braket{\psi}{\phi_{i_j}}\right|^2\\
    &=\frac{1}{\mcA_N}\int d\Omega_{N}\frac{1}{\mcN}\binom{N}{M}\brac{\left|\braket{\psi}{\phi_{i_1}}\right|^4+(M-1)\left|\braket{\psi}{\phi_{i_1}}\right|^2\left|\braket{\psi}{\phi_{i_2}}\right|^2},
  \end{split}
\end{equation}
where we have applied the isotropy property. This can be calculated directly using the representation (\ref{eq:2}) and the measure (\ref{eq:5}), and the result is
\begin{equation}
  \label{eq:36}
  \bar{f}_{st.est.}=\frac{1+1/M}{N+1}
\end{equation}
attaining the upper bound of Banaszek, \cite{art157}.

Based solely on the isotropic average over incident quantum states, we proved the relationship (\ref{eq:16}), (\ref{eq:30})
\begin{equation}
  \label{eq:47}
  F_{N\to M}=\frac{M-1}{N-1}+\frac{N-M}{N-1}F_{N\to 1}
\end{equation}
between the fidelity $F_{N\to M}$ of the desired task and the state estimation fidelity $F_{N\to 1}$. The state estimation fidelity is computed by an explicit integration over the state space. Our most general result is obtained in the case where the quantum state of interest is part of a pure state on an enlarged tensor product space $\mcH\otimes\mcH_R$, and thus our protocol is applied to teleport entanglement. The expression for $F_{N\to M}$ (\ref{eq:35}) shows that one may perform the operations in steps via states of intermediate dimensions $M < K < N$ without loss of fidelity $F_{N\to M}=F_{N\to K}F_{K\to M}$. 

Our protocol reaches the theoretical maximum fidelity for teleportation of pure states and pure entangled states, and for the chosen averaging over density matrices our protocol is the best among diagonal POVM's, but it is an open question if other POVM schemes can perform better for mixed states. Different values of $R$ lead to different distributions over the space of density matrices. $R=1$, for example, corresponds to pure states only, and $R\le N$ in general produces density matrices with rank less than or equal to $R$. Our explicit calculation of the $F_{N\to 1}$ fidelities lend themselves to analyses where different promises are given about the incident state, leading to a change in the integration measure $d\Omega_N$. One may assign prior probability measures, for example restrict the calculations to real Hilbert spaces. As long as the isotropy is maintained our general formula (\ref{eq:47}) holds. Other choices of ``uniform'' probability distributions on the space of mixed states might be of interest too, see Ref.~\cite{art291}.

This work was supported by The Danish Research Foundation -- Danmarks Grundforskningsfond. We thank Uffe V. Poulsen and Ole S\o rensen for useful discussions.

\appendix

\section{Integration measure on a complex Hilbert space}
\label{sec:jacobian}

We must determine the Jacobian required for changing between the two sets of complex cartesian and hyperspherical coordinates related by the transformation
\begin{equation}
  \label{eq:9}
    \brac{\begin{array}{c}
       z_1\\z_2\\z_3\\\vdots\\z_{N-1}\\z_N
    \end{array}}
    =\brac{\begin{array}{l}
       r\cos\theta_{1}e^{i\phi_{1}}\\
       r\sin\theta_{1}\cos\theta_{2}e^{i\phi_{2}}\\
       r\sin\theta_{1}\sin\theta_{2}\cos\theta_{3}e^{i\phi_{3}}\\
       \hspace{.5cm}\vdots\hspace{1cm}\vdots\hspace{1cm}\ddots\hspace{.5cm}\ddots\\
       r\sin\theta_{1}\sin\theta_{2}\cdots\sin\theta_{N-2}\cos\theta_{N-1}e^{i\phi_{N-1}}\\
       r\sin\theta_{1}\sin\theta_{2}\cdots\sin\theta_{N-2}\sin\theta_{N-1}e^{i\phi_{N}}
    \end{array}},
\hspace{.5cm}
\begin{array}{c}
0\le r < \infty\\
0\le\theta_{1},\ldots,\theta_{N-1}\le\frac{\pi}{2}\\
0\le\phi_{1},\ldots,\phi_{N}\le 2\pi
\end{array}.
\end{equation}

For real polar coordinates
\begin{equation}
  \label{eq:10}
  d(\rho\cos\alpha)d(\rho\sin\alpha)=\rho d\rho d\alpha,
\end{equation}
and hence for a complex $z_1=x_1+ix_2=r\cos\theta_{1}e^{i\phi_{1}}$ we have
\begin{equation}
  \label{eq:3}  
  dx_{1}dx_{2}=(r\cos\theta_{1})d(r\cos\theta_{1})d\phi_{1}.
\end{equation}
For $N=2$, with $z_1=x_1+ix_2$, $z_2=x_3+ix_4$ we have
\begin{equation}
  \label{eq:11}
  \begin{split}
    &dx_{1}dx_{2}dx_{3}dx_{4}=(dx_{1}dx_{2})(dx_{3}dx_{4})
    =(r\cos\theta_{1})d(r\cos\theta_{1})d\phi_{1}(r\sin\theta_{1})d(r\sin\theta_{1})d\phi_{2}\\
    &=r^2\cos\theta_{1}\sin\theta_{1}d(r\cos\theta_{1})d(r\sin\theta_{1})d\phi_{1}d\phi_{2}
    =r^3\cos\theta_{1}\sin\theta_{1}drd\theta_{1}d\phi_{1}d\phi_{2},
  \end{split}
\end{equation}
and hence the Jacobian for $N=2$ is
\begin{equation}
  \label{eq:12}
  \mcJ_{2}(r,\theta_{1})=r^3\cos\theta_{1}\sin\theta_{1}.
\end{equation}
For a general $N$
\begin{equation}
  \label{eq:13}
  \begin{split}
    &dx_{1}dx_{2}dx_{3}\cdots dx_{2N}=(dx_{1}dx_{2})(dx_{3}\cdots dx_{2N})\\
    &=(r\cos\theta_{1})d(r\cos\theta_{1})d\phi_{1}
    \mcJ_{N-1}(r\sin\theta_1,\theta_2,\theta_3,\ldots,\theta_{N-1})
    d(r\sin\theta_1)d\theta_2 d\theta_3\cdots d\theta_{N-1}
    d\phi_{2}\cdots d\phi_{N}\\
    &=r^2\cos\theta_{1}
    \mcJ_{N-1}(r\sin\theta_1,\theta_2,\theta_3,\ldots,\theta_{N-1})
    dr d\theta_{1}d\theta_{2}\cdots d\theta_{N-1}
    d\phi_{1}d\phi_{2}\cdots d\phi_{N},
  \end{split}
\end{equation}
and therefore
\begin{equation}
  \label{eq:14}
  \begin{split}
    &\mcJ_{N}(r,\theta_1,\theta_2,\ldots,\theta_{N-1})=
    r^2\cos\theta_{1}
    \mcJ_{N-1}(r\sin\theta_1,\theta_2,\theta_3,\ldots,\theta_{N-1})\\
    &=r^2\cos\theta_{1}(\sin\theta_1)^{2(N-1)-1}
    \mcJ_{N-1}(r,\theta_2,\theta_3,\ldots,\theta_{N-1})\\
    &=\brac{r^2\cos\theta_{1}\sin\theta_{1}(\sin^2\theta_{1})^{N-2}}
    \brac{r^2\cos\theta_{2}\sin\theta_{2}(\sin^2\theta_{2})^{N-3}}
    \mcJ_{N-2}(r,\theta_3,\theta_{4}\ldots,\theta_{N-1})\\
    &=\prod_{k=1}^{N-2}r^2\cos\theta_{k}\sin\theta_{k}(\sin^2\theta_{k})^{N-k-1}\mcJ_{2}(r,\theta_{N-1})
    =r^{2N-1}\prod_{k=1}^{N-1}\cos\theta_{k}\sin\theta_{k}(\sin^2\theta_{k})^{N-k-1},
  \end{split}
\end{equation}
where we have used equation~(\ref{eq:12}) for $\mcJ_2$, and the fact that $\mcJ_{k}(\rho,\cdot)\propto \rho^{2k-1}$.

Equation~(\ref{eq:5}) now follows by noting that
\begin{equation}
  \label{eq:15}
    d\Omega_{N}=\frac{dV_{N}}{dr}\Bigg\vert_{r=1}
    =\mcJ_{N}(1,\theta_{1},\theta_{2},\ldots,\theta_{N-1})d\theta_{1}d\theta_{2}\cdots d\theta_{N-1}d\phi_{1}d\phi_{2}\cdots d\phi_{N}.
\end{equation}
                                

\end{document}